# Properties of spin ½ triangular lattice antiferromagnets: Cu$RE_2$Ge$_2$O$_8$ ($RE$=Y, La)

Hwanbeom Cho[1,2], Marie Kratochvílová[1,2], Hasung Sim[1,2], Ki-Young Choi[1,2], Choong Hyun Kim[1,2], Carley Paulsen[3], Maxim Avdeev[4], Darren C. Peets[1,2], Younghun Jo[5], Sanghyun Lee[6], Yukio Noda[7], Michael J. Lawler[8], and Je-Geun Park[1,2#]

[1]Center for Correlated Electron Systems, Institute for Basic Science, Seoul 08826, Korea

[2]Department of Physics & Astronomy, Seoul National University, Seoul 08826, Korea

[3]Institut Néel, C.N.R.S-Université Grenoble Alps, BP 166, 38042 Grenoble, France

[4]Australian Nuclear Science and Technology Organisation, New Illawarra Road, Lucas Heights, New South Wales, 2234, Australia

[5]Quantum Materials Research Team, Korea Basic Science Institute, Daejeon 305-333, Korea

[6]Institute of Materials Structure Science, KEK, Tokai, Ibaraki, 319-1106, Japan

[7]Institute of Multidisciplinary Research for Advanced Materials, Tohoku University, Sendai 980-8577, Japan

[8]Department of physics, Binghamton University, Vestal NY 13850, USA

# Corresponding author: jgpark10@snu.ac.kr

**Abstract**: We found new two-dimensional (2D) quantum ($S$=1/2) antiferromagnetic systems: Cu$RE_2$Ge$_2$O$_8$ ($RE$=Y and La). According to our analysis of high-resolution X-ray and neutron diffraction experiments, the Cu-network of Cu$RE_2$Ge$_2$O$_8$ ($RE$=Y and La) exhibits a 2D triangular lattice linked via weak bonds along the perpendicular $b$-axis. Our bulk characterizations from 0.08 to 400 K show that they undergo a long-range order at 0.51(1) and 1.09(4) K for the Y and La systems, respectively. Interestingly, they also exhibit field induced phase transitions. For theoretical understanding, we carried out the density functional theory (DFT) band calculations to find that they are typical charge-transfer-type insulators with a gap of $E_g \cong 2$ eV. Taken together, our observations make Cu$RE_2$Ge$_2$O$_8$ ($RE$=Y and La) additional examples of low-dimensional quantum spin triangular antiferromagnets with the low-temperature magnetic ordering.

I. Introduction

A Heisenberg model is a convenient theoretical tool that we can use to understand the ground state of an isotropic magnetic system. When the exchange interaction between neighboring spins is of antiferromagnetic (AF) nature, two competing ground states capture the landscape of possibilities: a simple Néel-type long-range ordered (LRO) state described within the framework of classical mechanics [1] and a spin-singlet state formed by a superposition of spin states, which is a purely quantum mechanical concept [2]. The energy of the Néel-type ground state is $-\frac{z}{2}S^2J$ per spin with $z$ being the number of the nearest neighbors. This ground state energy varies depending on one-dimensional (1D) chain, two-dimensional (2D) square lattice and three-dimensional (3D) cubic systems. On the other hand, the energy per spin of the singlet ground state is $-\frac{1}{2}S(S+1)J$ for all three cases. Dimension therefore naturally controls the competition.

Specifically, the singlet state is a competitive ground state for 1D quantum spin ($S = \frac{1}{2}$) systems because of its lower energy whereas the Néel-ordered state wins over the singlet ground state for 3D systems. For 2D systems, however, there is competition between the Néel-ordered state and the singlet state since the energies of both cases become very similar to one another. This argument makes the 2D magnetic systems interesting and explains why, as compared with either 1D or 3D systems, 2D systems have a tendency of singlet and Néel-ordered ground states found in close proximity in the parameter space of physical quantities [3-14]. In addition, geometrical frustration can further prevent a system from ordering in either simple Néel type or singlet type states and thus can lead to exotic physical phenomena at low temperatures [6-18]. This is especially true for 2D triangular lattice systems as compared with 2D square lattice systems, the latter of which is often found in the Néel-ordered state due to the weakness of the diagonal next-nearest neighbor interactions [19].

Remarkably, in 2D triangular Heisenberg antiferromagnet (HAF) systems, quantum mechanics can do its work in another way. For 2D HAF systems that enter a LRO ground state, various spin configurations appear suddenly under magnetic fields [6-12,15-18]. For classical spins, the ground state is expected to be the so-called 120º structure with the ground state being continuously degenerate [20]. As an external magnetic field is applied and increases in strength, spins become preferentially aligned along the field direction. Therefore, one would naturally expect a monotonous magnetic field dependence of the magnetization curve $M(H)$ [16,20,21]. However, in real quantum 2D triangular systems the combination of low dimensionality, geometric frustration, and quantum spin nature remarkably enhances quantum fluctuations that break the otherwise degenerate ground state. Consequently, there are sometimes abrupt changes between different spin configurations leading to a plateau-type behavior in the $M(H)$ curves [15-18].

For example, a 1/3 plateau phenomenon was observed some time ago in $Cs_2CuBr_4$, a system regarded as a quasi-2D HAF

system [6] and again more recently in $Ba_3CoSb_2O_9$, a system regarded as an ideal 2D HAF system [7]. In both cases, the 120º ordered spin configuration changes to an up-up-down (*uud*) configuration, inducing the emergence of the plateau with one-third of the saturated magnetization, $M_s$ in the $M(H)$ curve. The magnetic structure further changes into a coplanar configuration at higher magnetic fields. Moreover, other anomalies in $M(H)$ have been reported at higher fractions of $M_s$ that can originate from other competing quantum spin phases. Notable examples of such phases include those at 1/2, 5/9 and 2/3 of $M_s$ in $Cs_2CuBr_4$ [8] and a phase with 3/5 of $M_s$ in $Ba_3CoSb_2O_9$, [9-11]. Interestingly, $Cs_2CuCl_4$, isostructural with $Cs_2CuBr_4$, is reported to have a fractional quantum spin-liquid phase above the transition temperature of the LRO ground state [14]. On the other hand, in $CsCuCl_3$, another quasi-2D HAF system, a quantum phase transition was previously identified at a transition point from an umbrella-like configuration to a coplanar configuration when the magnetic field is applied along the *c*-axis. This quantum phase transition originates from the competition between quantum fluctuations and a magnetic anisotropic energy, which yields a step-like behavior in the $M(H)$ curve [12].

As demonstrated so many times in the field before, finding a new sample has been one of the key driving forces in pushing the boundary of quantum magnetism and so unearthing new phenomenon. In particular, materials with a new lattice have been a fertile ground for finding novel physics. Here, we identify a possibly new form of 2D quantum magnetism in Cu*RE*$_2$Ge$_2$O$_8$ (*RE* = Y and La). For example, one earlier structural work reported that CuNd$_2$Ge$_2$O$_8$ system (*C* 1 *m* 1, monoclinic) has a quasi-1D copper octahedral (CuO$_6$) corner-shared zigzag chain along the *a*-axis while neodymium oxide dodecahedral (NdO$_8$) co-linked chains are running along the *c*-axis [22]. Spectroscopic measurements were also performed to investigate various types of germanium oxide polyhedra by classifying vibration modes [22,23]. In these previous works, the $Nd^{3+}$ ion (total angular momentum, $J = 9/2$) plays a dominant role in determining the physical properties of CuNd$_2$Ge$_2$O$_8$ system. Unfortunately, this dominant role played by Nd then makes it harder to study the otherwise interesting physical phenomena arising from the 1D copper quantum spins. For instance, magnetic measurements have shown simple paramagnetic behavior with an anomaly attributed to the crystal-field splitting of Nd ions [22]. As far as we are aware, there has been no single crystal studies on the same chemical formula with non-magnetic rare-earth element such as Y and La. Furthermore, as we demonstrate below it is very important for us to determine the chemical structure accurately using a single crystal sample in order to understand better the physical properties.

In this paper, we report a comprehensive study of the crystal structure and physical properties of Cu*RE*$_2$Ge$_2$O$_8$ (*RE* = Y, La) by using both single crystal and polycrystalline samples. By choosing a non-magnetic version of the Cu*RE*$_2$Ge$_2$O$_8$ system with Y and La, we aim to study the physical phenomena solely related to the quantum spin of $Cu^{2+}$ and want to explore its behavior in the new low-dimensional lattice. By carrying out bulk measurements down to 0.08 K and up to 90 kOe, in addition to high-resolution single crystal and powder diffraction measurements, we demonstrate that the Cu*RE*$_2$Ge$_2$O$_8$ system is a new quantum magnetic system revealing a 2D triangular lattice and novel magnetic field-induced phase transitions.

II. Experimental methods

We synthesized single crystals of $CuLa_2Ge_2O_8$ using a flux method and polycrystals of $CuY_2Ge_2O_8$ and $CuLa_2Ge_2O_8$ via a solid state reaction method [22]. Our structure analysis using both x-ray and neutrons confirm that all our samples form in the space group $I\,1\,m\,1$ (see Figure 1(a)). To synthesize the single crystal, we used CuO and $GeO_2$ powder as self-flux so that the ratio of the initial materials was $CuO : La_2O_3 : GeO_2 = 8 : 1 : 6$. The mixture was annealed for 30 minutes at 1260℃ in a platinum crucible and then slowly cooled down to 1000℃ at a rate of 2℃/hour. We used diluted HCl (~17％) solution to remove from the grown crystals any residual flux and byproducts like CuO and $CuGeO_3$ [22]. The typical size of a $CuLa_2Ge_2O_8$ single crystals is about 0.001 $mm^3$ and a photo of one of the biggest crystals is shown in the inset of Figure 1(b). The color of the crystals is transparent blue and they have an albite shape similar to the one reported in Ref. [22]. A passing note, despite several attempts with the same flux method we failed to grow single crystals of $CuY_2Ge_2O_8$, which may well be due to the difference in the ionic size between Y and La. We also prepared high-quality polycrystalline samples of $CuY_2Ge_2O_8$ and $CuLa_2Ge_2O_8$ by sintering the mixture of CuO, $RE_2O_3$, and $GeO_2$ in a stoichiometric ratio, 1 : 1 : 2 at 850℃ for 12 hours. The mixture was ground before a final sintering under the same conditions. The finally sintered powder of $CuY_2Ge_2O_8$ ($CuLa_2Ge_2O_8$) was then pelletized and annealed by increasing temperature from 950℃ to 1100 (1050)℃ at 50℃ steps with a duration of 24 hours at each step: we optimized the final sintering temperature to produce highest-quality samples based on our X-ray diffraction (XRD) data.

To check the quality of the samples we used three different diffraction instruments: a high-resolution single crystal X-ray diffractometer (XtaLAB P200, Rigaku), a powder X-ray diffractometer (Miniflex II, Rigaku) and a high-resolution neutron diffractometer (Echidna, ANSTO). For the full structural refinement of $CuLa_2Ge_2O_8$, we mainly used XtaLAB P200, Rigaku with a wavelength of 0.710747 Å (a Mo target, averaged $K_α$). For the structural refinement of $CuY_2Ge_2O_8$ we used two powder diffractometers: Miniflex II, Rigaku with a wavelength of 1.540590 Å and 1.544310 Å ($K_{α1}$ and $K_{α2}$ respectively) and Echidna, ANSTO with a wavelength of 2.4378 Å. All analysis was done using the FullProf [24] software.

To orient single crystals for bulk measurements, we used an Imaging Plate XRD Laue Camera (IP-XRD, IPX Co., Ltd.). For example, a photo in Figure 1(b) shows a single crystal with the *b*-axis of the sample perpendicular to the flat plane with an area of 750×374 $μm^2$: this particular sample was used for the magnetization measurements. Magnetic and thermodynamic properties of polycrystalline samples were characterized from 0.4 to 350 K using MPMS-XL5 and PPMS-9ECII (Quantum Design) equipped with a $^3$He option. We used an MPMS-3 (Quantum Design) to measure the magnetization of the small single-crystal samples for the La case as well. For further low-temperature measurements of CuY$_2$Ge$_2$O$_8$, we employed a home-made SQUID magnetometer equipped with a dilution refrigerator with a base temperature of 84 mK. Finally, in order to understand the ground-state properties of CuLa$_2$Ge$_2$O$_8$ we carried out the electronic structure calculations and the density of states (DOS) using the Local Density Approximation with the Hubbard $U$ (LDA+$U$) density functional theory (DFT) with the software Vienna Ab initio Simulation Package (VASP) [25,26].

III. Experimental results and Analysis

1. Structural analysis

For the full structural refinement of CuLa$_2$Ge$_2$O$_8$, we used a small and high-quality single crystal with the dimension of 95×70×50 μm$^3$ to minimize the X-ray absorption effects. Using the high-resolution single crystal X-ray diffractometer, we collected 3383 Bragg peaks (2369 independent Bragg peaks) and refined the crystal structure. For the structural refinement of CuY$_2$Ge$_2$O$_8$, we also used data obtained from two powder diffractometers: a powder X-ray diffractometer and a neutron powder diffractometer with the aid of the structural information we obtained from the single crystal of CuLa$_2$Ge$_2$O$_8$.

The refined crystal structure of Cu$RE_2$Ge$_2$O$_8$ ($RE$ = Y, La) system is found to have a 2D triangular magnetic lattice as shown in Figure 2, where the blue, green and grey balls represent the Cu, La and Ge atoms, respectively. Oxygen atoms are omitted for clarity and instead their positions are indicated by the red ends of sticks stretched from the copper atoms to the corner of the germanium oxide polyhedra. The final structures also show the CuO$_6$ corner-shared chains running along the $a$-axis by linking the copper and oxygen atoms as in the CuNd$_2$Ge$_2$O$_8$ case [22] with the $C$ 1 $m$ 1 unit cell setting (for the $I$ 1 $m$ 1 unit cell setting the chain direction runs along the $J_3$ direction as shown in Figure 2(b)). However, there is one striking difference between our refined final structure and that reported of CuNd$_2$Ge$_2$O$_8$ [22]; which is that according to our analysis the copper octahedrons are highly distorted due to the Jahn-Teller effect while the earlier work shows a significantly smaller distortion. For example, in our structural model the bond lengths between the central copper and oxygen atoms on the distorted equatorial plane (O6~O9) are about 1.76(1)~1.98(1) Å whereas the bond lengths between the copper and two apical oxygen atoms (O10) are 2.63(2) and 2.71(1) Å, respectively (see Figure 2(c)). As the apical bond lengths are larger than 2.13 Å, which is the sum of ionic radii of Cu$^{2+}$ and O$^{2-}$, the orbital overlap between Cu$^{2+}$ and O$^{2-}$ is expected to be very small along the apical direction. Therefore, we conclude that the copper oxide polyhedra of Cu$RE_2$Ge$_2$O$_8$ ($RE$ = Y, La) does not have the usual CuO$_6$ octahedron, which is rather unusual among Cu oxides. Instead, we consider the CuO$_4$ plaquette as a basic building block. This then leads us to rethink about the crystal structure. After trials with several other possibilities, we conclude that in contrast to the previous claim [22] for CuNd$_2$Ge$_2$O$_8$ there is no evidence of the 1D chain in our final refined crystal structure of CuLa$_2$Ge$_2$O$_8$. Since there is no 1D chain along the $a$-axis, we think that there is no reason for us to stick with the *C* 1 *m* 1 unit cell setting used in Ref. [22]. Thus, instead we adopted the *I* 1 *m* 1 cell setting in our structural analysis, which is more appropriate for the description of the coordination of the copper ion by transforming *c* and *a+c* vector of *C* 1 *m* 1 to *a* and *c* unit vector of *I* 1 *m* 1 respectively. Typical refinement results are shown in Figure 1. Table 1 summarizes the structural information of the lattice parameters, atomic positions, thermal parameters and agreement factors for CuLa$_2$Ge$_2$O$_8$.

For the benefit of our later discussion on the bulk properties, and in particular the magnetic properties, let us examine the crystal structure further in detail. The distorted CuO$_4$ plaquette can interact with the adjacent ones so that we can assign four different exchange integrals: $J_1$, $J_2$, $J_3$ and $J_4$ (see Figure 2(a)). The direct distances between the neighboring copper ions along each path are 5.16(1)~5.21(1) Å ($J_1$, $J_2$, $J_3$) and 6.48(2) Å ($J_4$), respectively. Therefore, the exchange interactions along $J_1$, $J_2$ and $J_3$ are expected to be about the same while $J_4$ should be smaller than the other three. In this scenario, the magnetic lattice of Cu ions in the Cu*RE*$_2$Ge$_2$O$_8$ (*RE* = Y, La) system becomes a 2D triangular lattice within the *ac* plane (see Figure 2(b)). At the same time, we comment that the actual magnitude of the exchange interaction is expected to be significantly smaller than that of other Cu oxides including cuprates [27~29] for the following reason. The neighboring CuO$_4$ plaquettes are well separated by germanium and RE atoms; Ge1 and Ge2 atoms make GeO$_4$ tetrahedrons that form a grid on the *ac* plane separating copper oxide triangular plane (see Figure 2(a)) whereas Ge3 atoms construct GeO$_5$ trigonal bipyramids that block the copper-oxygen-copper link in the triangle (see Figure 2(c)).

2. Magnetic & heat capacity measurements

We measured the temperature-dependent direct current (DC) magnetic susceptibility using the polycrystalline samples. Two anomalies are immediately obvious in the data shown in Figure 3(a): a broad hump centered at 0.89(1) and 1.40(4) K and an inflection point (indicated by an arrow) at $T_{\text{LRO}}$ = 0.50(1) and 0.93(2) K for Y and La, respectively. The broad maximum is usually a characteristic sign of low-dimensional magnetic systems due to the AF short-range interaction [30-38]. In Figure 3(b), we can clearly see a peak in the temperature derivative of the susceptibility curve at the same temperature as the inflection point in the susceptibility. This peak then suggests an onset of the LRO state and will be discussed together with the thermodynamic data later in the paper.

From the Curie-Weiss fit (see the insert of Fig. 3a), we obtained the same effective moment value of $\mu_{\text{eff}}$ = 1.94(1)$\mu_B$/f.u. for both Cu*RE*$_2$Ge$_2$O$_8$ (*RE* = Y, La) and the Curie-Weiss temperature $\theta_{\text{CW}}$ = -1.1(1) and -5.7(4) K for the Y and La compounds. We used the following formula, $\chi(T) = \chi_0 + C/(T - \theta_{\text{CW}})$ to fit the data over the temperature region of 6.8 - 350 K and 19.7 - 398 K for the Y and La compounds, respectively. In the equation, $\chi_0$ represents the temperature-independent Van Vleck paramagnetism from elements other than Cu ($\chi_0 \sim 2\times10^{-5}$ emu/mole·Oe for both cases) and $C$ is the Curie constant of the Cu spins. We note that the effective moment obtained from the fit is comparable with the theoretical effective moment of $S=1/2$ for $Cu^{2+}$ ($g_J\mu_B(S(S+1))^{1/2}$ =1.73$\mu_B$), supporting the scenario of the quantum spin system being realized in $CuRE_2Ge_2O_8$ ($RE$ = Y, La). The negative value of $\theta_{CW}$ shows the averaged exchange interaction to be of antiferromgnetic origin. The frustration factor $f = \frac{\theta_{CW}}{T_{LRO}}$ is found to be larger than unity for both cases ($f_{La}$=6.1(4) & $f_Y$=2.2(2)). This could have an origin in the effects of the geometrical frustration in the 2D triangular AF ground state [39]. We also comment that the average value of the exchange interaction estimated from the transition temperature is smaller by 2 – 3 orders of magnitude than the well-known cuprate systems [27,28,40,41].

Figures 3(c) and (d) present the $M(H)$ curves measured on the polycrystalline samples at various temperatures around $T_{LRO}$ and the insets show the first derivative of $M(H)$. The magnetization and its first derivative show typical behavior of a low-dimensional quantum spin system and magnetic field-induced phase transitions. For example, the overall shape of the $M(H)$ curve is convex rather than a straight line, which is consistent with the behavior of low-dimensional quantum systems [6,15,21,32,42]. In the inset of Figure 3(c) and (d) for both Y and La compounds, the d$M$/d$H$ curve shows paramagnetic behavior above $T_{LRO}$: the slope decreases gradually to zero as the external magnetic field approaches the saturation field. As the temperature decreases below $T_{LRO}$, a new field-induced phase transition appears. We note that around zero field the slope gets reduced as compared the $T > T_{LRO}$ case. As the external field increases, $M(H)$ displays an abrupt change at a low critical field $H_C$ (2.5(3) and 4.0(3) kOe for the Y and La compounds). Of further interest, the $M(H)$ curve shows a kink at $H_C$ for both compounds and there is no hysteresis around $H_C$ at the base temperature, indicative of a second order phase transition. After the abrupt change, the slope gradually increases with the magnetic field and finally reaches a maximum before decreasing to zero with the magnetization value approaching its saturation moment value. The saturation field of both compounds (14.0(3) and 38.0(3) kOe for Y and La compounds, respectively) is also comparable with the energy scales of the exchange interactions estimated from $\theta_{CW}$. We comment that the saturated magnetization value is close to 1$\mu_B$/$Cu^{2+}$, corresponding to a quantum spin system where $gS$ = 1 ($g$ = 2 is the Landé g-factor and $S$ = ½ is the spin angular momentum number).

In order to examine the field-induced transitions more closely, we measured $M(H)$ with magnetic field applied along two perpendicular directions using the single crystal of $CuLa_2Ge_2O_8$ as shown in Figure 1(b). Upon increasing the magnetic field along the $b$-axis perpendicular to the triangular plane, $M(H)$ increases but its derivative decreases until $H_{C1}$ = 0.80(5) kOe, where the slope becomes constant (see Figure 3(e)). This behavior at very low fields is difficult to see in the data obtained from the polycrystalline samples. When the magnetic field reaches $H_{C2}$ = 5.8(1) kOe, the d$M$/d$H$ starts to increase abruptly in a qualitatively similar manner to the polycrystalline $M(H)$ data and the magnetization keeps increasing with magnetic field. The second anomaly, $H_{C2}$ = 5.8(1) kOe corresponds to the critical point $H_C \approx$ 4.0(3) kOe observed in the polycrystalline samples; the difference between these two values could come from slightly different anisotropy values of the two samples measured. We note that there is a small difference in behavior in M(H) between the polycrystalline and single crystal samples. We think that this difference is likely to be due to the fact that some defects present in the polycrystalline sample either mask or weaken the transitions seen in the single crystalline sample. When the external magnetic field is applied perpendicular to the *b*-axis in the in-plane direction (see Figure 3(f)), the overall behavior of d$M$/d$H$ is similar to that of the polycrystalline case and also to the behavior for $H \parallel b$, except for the fact that $H_{C1}$ and $H_{C2}$ are too close to be distinguishable in the data for $H \perp b$. Because the overall shape of $M(H)$ for both crystallographic orientations is similar, the anisotropy seems to be small.

The heat capacity data of both the Y and La samples show a lambda-like sharp peak at $T_N = 0.51(1)$ and $1.09(4)$ K, respectively (see Figure 4(a)). These values are consistent with $T_{LRO}$ obtained from the magnetization measurement, which confirms further that both systems have a magnetically ordered ground state at low temperatures. In order to calculate the magnetic entropy change below the transition temperature we subtracted the phonon contribution from the raw data by using a Debye temperature of 381(1) and 332(2) K for the Y and La samples, respectively (see the inset in Figure 4(a)). The estimated magnetic entropy ($S_Y = 5.83(8)$ & $S_{La} = 5.52(3)$ J/mole·K) is comparable to the theoretical entropy for a quantum spin system ($R\ln2 = 5.763$ J/mole·K). In order to examine the field dependence of the phase transition, we also measured the field-dependent heat capacity of $CuLa_2Ge_2O_8$, which has a relatively higher transition temperature. As shown in Figure 4(b), the peak in the heat capacity gets substantially suppressed upon the application of magnetic field and eventually becomes lower than the base temperature of our $^3$He set-up. Looking at the phase diagram of Figure 4(d), we think that the transition temperature gets suppressed down to the absolute zero temperature at around 50 kOe. Our measurements on $CuY_2Ge_2O_8$ show similar behavior (see the inset of Figure 4(b)) but in this case, $T_N$ goes down to zero at around 15 kOe (see the solid line in Figure 4(c)). A passing comment, our observation of the sharp peak in the heat capacity (see Figure 4(a)) is at variance with what is expected from the Bonner and Fisher 1D HAF model [32]. Thus it further supports our view that the magnetic lattice of the Cu ions of $CuRE_2Ge_2O_8$ ($RE$ = Y and La) is more of a 2D triangular lattice, not a 1D HAF chain system.

3. DFT calculation

In order to gain further insights into the ground state properties and, in particular, the density of states of $CuLa_2Ge_2O_8$, we performed the first-principles density functional theory calculations using a LDA+$U$ method with the Perdew-Burke-Ernzerhof exchange-correlation density functional (PBEsol) form of exchange correlation functional as implemented in the VASP [25,26] program. We used a local Hubbard interaction of $U$=5 eV for Cu 3$d$ electrons, which gives a reasonable value of a band gap of about 2 eV in the density of states (DOS) (see Figure 5(a)). We note that this value is consistent with the blue color of our single crystal samples (see Figure 1(b)).

A comparison of the total DOS with the partial DOS from the copper *d*-orbitals in Figure 5(b) shows that the main contribution to the bands comes from the copper *d*-orbitals. However, our calculations also demonstrate that the germanium and lanthanum atoms contribute significantly to the DOS of the conduction band while the oxygen atoms contribute more to the DOS of the valence band. As a result, the valence band is mainly hybridized oxygen *p*-bands and copper *d*-bands. Therefore, this system can be considered as a charge-transfer-type insulator like many cuprates [43,44].

IV. Discussion and Conclusion

If we take the crystal structure, physical properties and DFT calculation results together, they suggest that the Cu$RE_2$Ge$_2$O$_8$ (*RE* = Y, La) system is a new quantum 2D triangular charge-transfer insulator. They exhibit the clear sign of the long-range order at low temperatures and the magnetic field-induced phase transitions at a low field range. The LRO is probably suppressed due to the combined effects of quantum spins, geometrical frustration, and low-dimensionality. To illustrate the uniqueness of this Cu$RE_2$Ge$_2$O$_8$ system, we would like to compare our physical properties with those of other known quantum 2D triangular systems in the following paragraphs. There are only few inorganic quantum 2D triangular AF systems, which have LRO ground states: CsCuCl$_3$[45,46], CsCuBr$_4$ (CsCuCl$_4$) [13,47~50] and Ba$_3$CoSb$_2$O$_9$ [7,9,51].

Compared with the three other systems, Cu$RE_2$Ge$_2$O$_8$ reveals a CuO$_4$ plaquette as a magnetic unit forming a 2D triangular lattice within the *ac* plane with the planes stacking along the *b*-axis. In Figure. 6, we compare the crystal structure of Cu$RE_2$Ge$_2$O$_8$ together with those of the three other compounds. In the *ac* plane of Cu$RE_2$Ge$_2$O$_8$, the distances between adjacent copper ions are almost the same ($J_1 \approx J_2 \approx J_3$) and they are shorter by 1.3(1) Å than the interlayer distance ($J_4$). However, unlike the other 2D triangular systems having triangular lattices on a relatively flat plane, in Cu$RE_2$Ge$_2$O$_8$ the triangular plane corrugates in a zig-zag shape by 146.19(4) degrees along the *a*-axis (see Figure 2(a)). This then is likely to introduce a larger value of Dzyaloshinskii-Moriya interaction, which would then have quite significant effect on the magnetic structure as well as spin dynamics as seen in many other systems such as multiferroic BiFeO$_3$ [52-55]. This corrugated plane and the mirror symmetry perpendicular to the *b*-axis suggest two types of the interlayer distances; along the *b*-axis the interlayer distance alternates between short and long ones, giving the unique structure for Cu$RE_2$Ge$_2$O$_8$. From this lattice structure, we expect that the system can have an exotic magnetic lattice with a strong and regular triangular interaction within the *ac* plane. As a result, the average exchange interaction is of an AF type and a weak interlayer interaction is formed along the *b*-axis alternatively.

It is rather common that 2D triangular AF systems show jump-like transitions in their M(H) data: CsCuCl$_3$ [56]; Cs$_2$CuBr$_4$ and Ba$_3$CoSb$_2$O$_9$ [6-11]. An exception is Cs$_2$CuCl$_4$ that does not show the $\frac{1}{3}M_s$ plateau [57]. In comparison, CuLa$_2$Ge$_2$O$_8$ shows two magnetic field-induced phase transitions at $H_{C1}$ = 0.80(5) kOe and $H_{C2}$ = 5.8(1) kOe (in the Y-based compound, $H_C$ = 2.5(3) kOe). As the slope between two critical points is constant, the spin state at this region seems to sustain its configuration similarly to the plateau like behavior of the *uud* state in a regular triangular lattice. However, the fractions of M(H) to $M_s$ are 0.019 and 0.117 at $H_{C1}$ = 0.80(5) kOe and $H_{C2}$ = 5.8(1) kOe, respectively: $M(H_C)/M_s$ = 0.084 for the Y-based compound. To our best knowledge, such small fraction numbers have not been reported before for a 2D triangular system. To understand further the field-induced phases, we need neutron diffraction studies at low-temperature.

To summarize, we have carried out the extensive study of the bulk properties of Cu*RE*$_2$Ge$_2$O$_8$ (*RE* = Y, La) including a crystal structure analysis, magnetic and thermodynamic measurements, and DFT calculations. Our results taken together suggest that the system should be a charge-transfer type quantum 2D triangular antiferromagnet with the clear sign of an antiferromagnetic transition at $T_N$ = 0.51(1) and 1.09(4) K for Y and La, respectively. Remarkably, a field-tuned phase transition arises between this magnetic order and a plateau phase with a small magnetization fraction.

**Acknowledgments**

We should acknowledge the Applications Group (Tom Hogan and Randy K. Dumas) at Quantum Design for collecting some of the low-temperature heat capacity data for this work. We would also like to thank Joosung Oh and Daniel Khomskii for useful discussions. We acknowledge the support of the Bragg Institute, Australian Nuclear Science and Technology Organisation, in providing neutron research facilities used in this work. We also acknowledge Soonmin Kang and Nahyun Lee for their technical assistance. This work was supported by Institute for Basic Science (IBS-R009-G1).

**Figure Captions**

Figure 1. (Color online) (a) The refinement results of the X-ray powder diffraction of Cu$RE_2$Ge$_2$O$_8$ ($RE$ = Y and La) and neutron powder diffraction data of CuY$_2$Ge$_2$O$_8$ measured in Echidna, ANSTO. (b) The refinement results of the single-crystal XRD data of CuLa$_2$Ge$_2$O$_8$ and a picture of single crystal sample taken using an optical microscope (in the inset).

Figure 2. Crystal structure of CuLa$_2$Ge$_2$O$_8$ in the $I\,1\,m\,1$ unit cell setting. The blue, green and grey balls indicate the copper, lanthanum and germanium atoms, respectively. (a) There are four different exchange paths: $J_1$, $J_2$, $J_3$ and $J_4$. Ge1 and Ge2 atoms make GeO$_4$ tetrahedrons that form a grid on the $ac$ plane and Ge3 atoms construct GeO$_5$ trigonal bipyramids. (b) CuO$_4$ plaquette forms a 2D triangular lattice within the $ac$ plane. (c) The Cu-O10 lengths are much longer than Cu-O6~O9 lengths on the distorted equatorial plane while La and Ge atoms are separating the nearest CuO$_4$ plaquettes and so prohibiting the formation of the Cu-O-Cu bonding between the adjacent CuO$_4$ plaquettes.

Figure 3. (Color online) (a) Temperature-dependent DC magnetic susceptibility $\chi_m$ of Cu$RE_2$Ge$_2$O$_8$ ($RE$ = Y and La). The arrows indicate the inflection points originating from a long-range order. The inset shows the inverse susceptibility curves and the Curie-Weiss fitting results (dashed lines). (b) Temperature derivative of the magnetic susceptibility of Cu$RE_2$Ge$_2$O$_8$ ($RE$ = Y and La). (c) Field-dependent magnetization $M(H)$ at various temperatures using polycrystalline CuLa$_2$Ge$_2$O$_8$. The first derivatives for both systems are shown in the inset. The position of the magnetic field-induced phase transition is indicated by an arrow. (d) Field-dependent magnetization $M(H)$ for polycrystalline CuY$_2$Ge$_2$O$_8$ and its first derivative are shown in the inset. The position of the magnetic field-induced phase transition is indicated by an arrow. (e) Magnetization $M(H)$ in $H \parallel b$ for the CuLa$_2$Ge$_2$O$_8$ single crystal at various temperatures and the first derivatives of $M(H)$ are shown in the inset. (f) Comparison of d$M$/d$H$ of the polycrystalline sample and single crystal in the crystallographic orientations $H \parallel b$ and $H \perp b$, respectively.

Figure 4. (Color online) Heat capacity $C_p(T)$ of Cu$RE_2$Ge$_2$O$_8$ ($RE$ = Y and La). (a) There is a lambda-like sharp peak in both systems indicating a long-range order. The solid and dashed lines are fits for the $T^3$-law to remove the phonon contribution. The inset shows the temperature dependence of magnetic heat capacity $C_m$ divided by temperature. (b) The field-dependent heat capacity measured data on CuY$_2$Ge$_2$O$_8$ (inset) and CuLa$_2$Ge$_2$O$_8$, which show the suppression of the long-ranged order with the external magnetic field. (c) and (d) The field-dependent transition points for the Y and La samples, respectively. The black crosses represent the Néel

temperatures ($T_N$) at each fields and the black diamond shows the saturation field ($H_S$) from the magnetization data. The solid line is the guide to eyes, indicating the existence of a critical field value, where the magnetic phase transition gets suppressed to the absolute zero temperature.

Figure 5. (Color online) (a) The total density of states calculated using the crystal structure of CuLa$_2$Ge$_2$O$_8$ with the Coulomb interaction, $U$ = 5 eV. (b) The partial density of states projected on the copper $d$-orbitals.

Figure 6. (Color online) Diagrams of magnetic lattices of selected inorganic quantum 2D triangular systems. The solid arrows depict the strong exchange interactions and the dashed ones depict the weak exchange interactions. CsCuCl$_3$ has strong FM interlayer ($J_\perp$) interactions but weak AF triangular interactions ($J$, $J´$)) whereas Cs$_2$CuBr$_4$ and Cs$_2$CuCl$_4$ have strong AF chain interactions ($J$) but weak AF inter-chain interactions ($J´$) and weak interlayer interactions ($J_\perp$). Ba$_3$CoSb$_2$O$_9$ shows a strong and regular AF triangular interaction ($J$, $J´$) but weak AF interlayer interaction ($J_\perp$). On the other hand, Cu$RE$$_2$Ge$_2$O$_8$ ($RE$ = Y and La) has a strong and regular AF triangular interaction ($J_1 \cong J_2 \cong J_3$) but weak interlayer interaction ($J_4$) formed alternatively along the stacking axis.

**Figure 1**

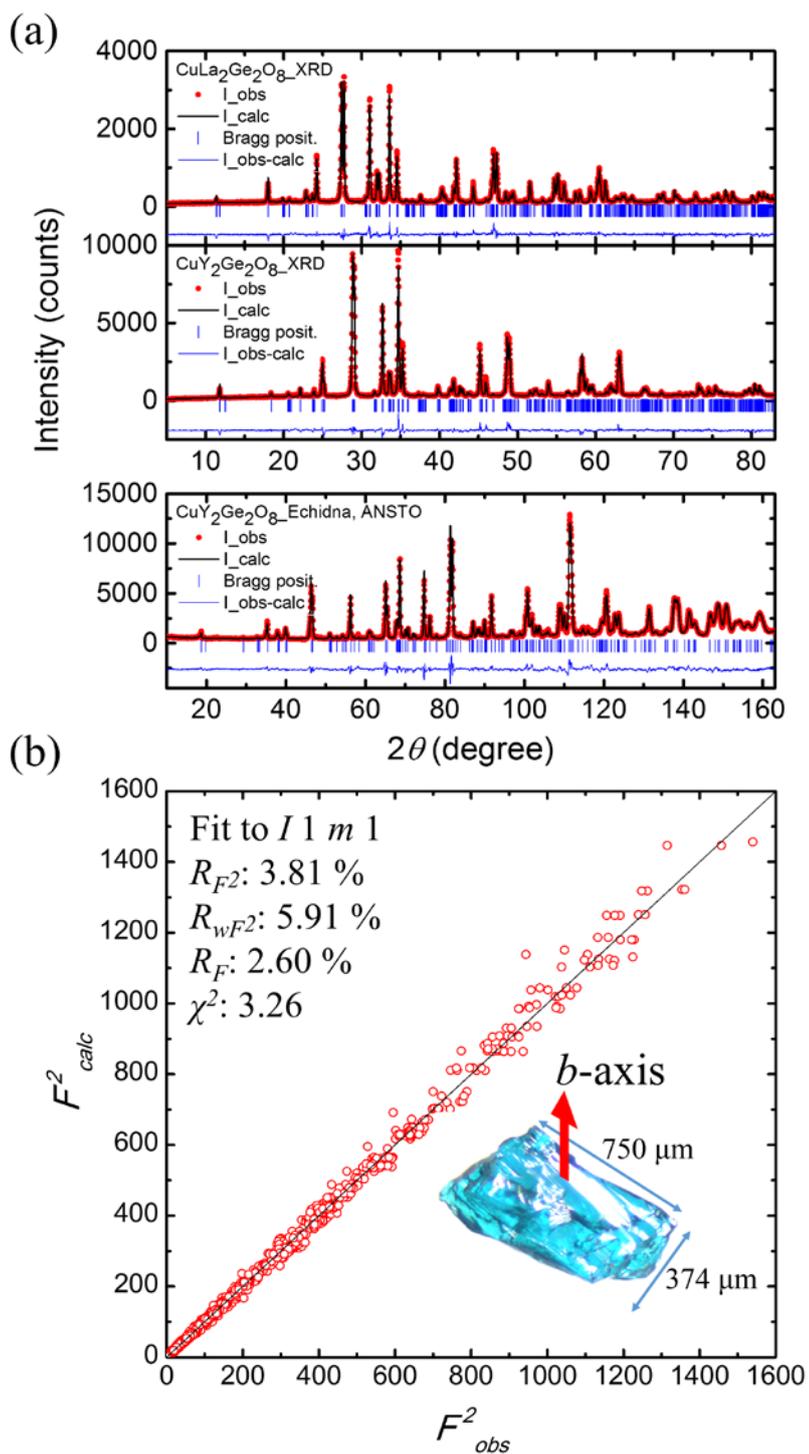

**Figure 2**

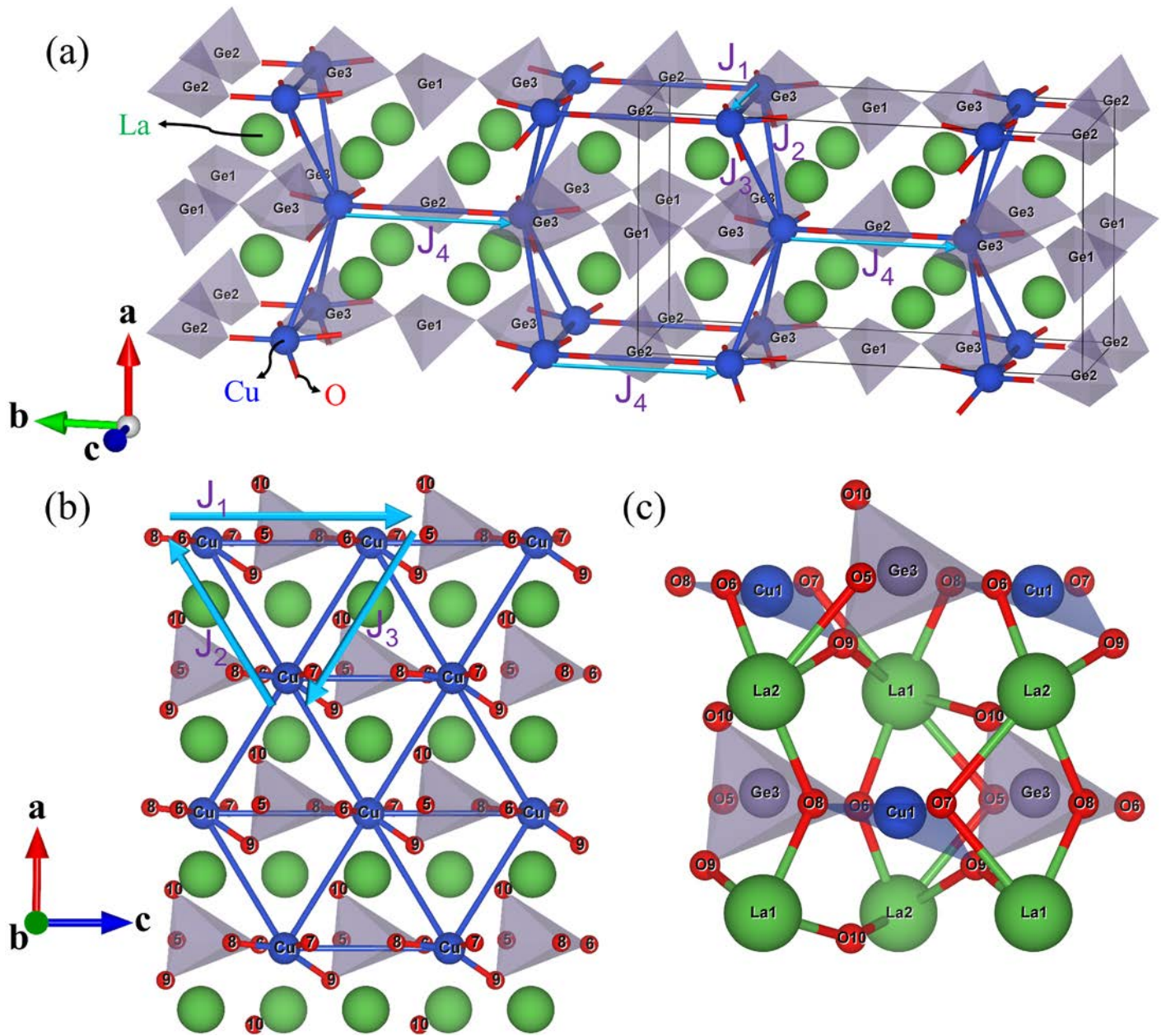

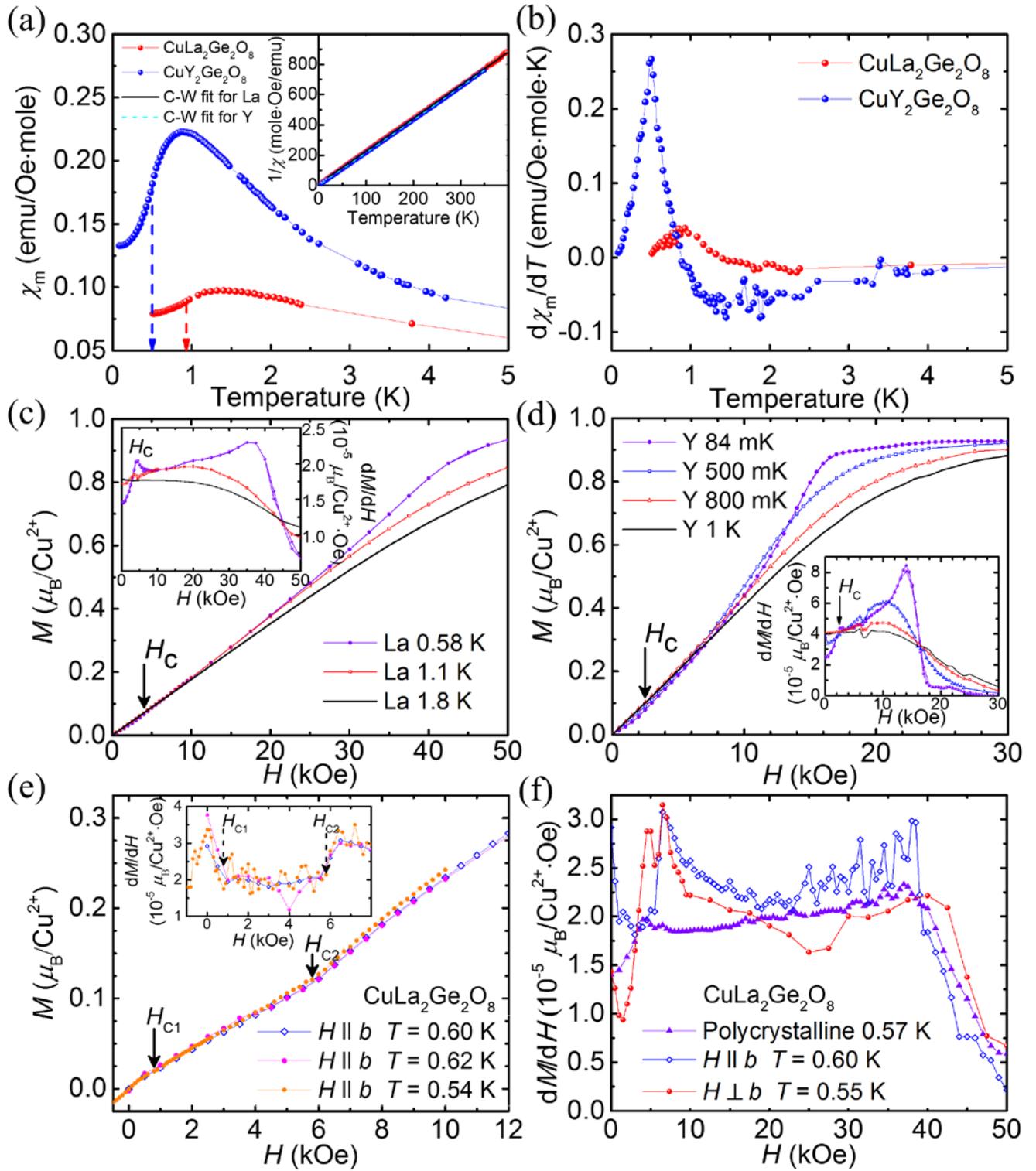

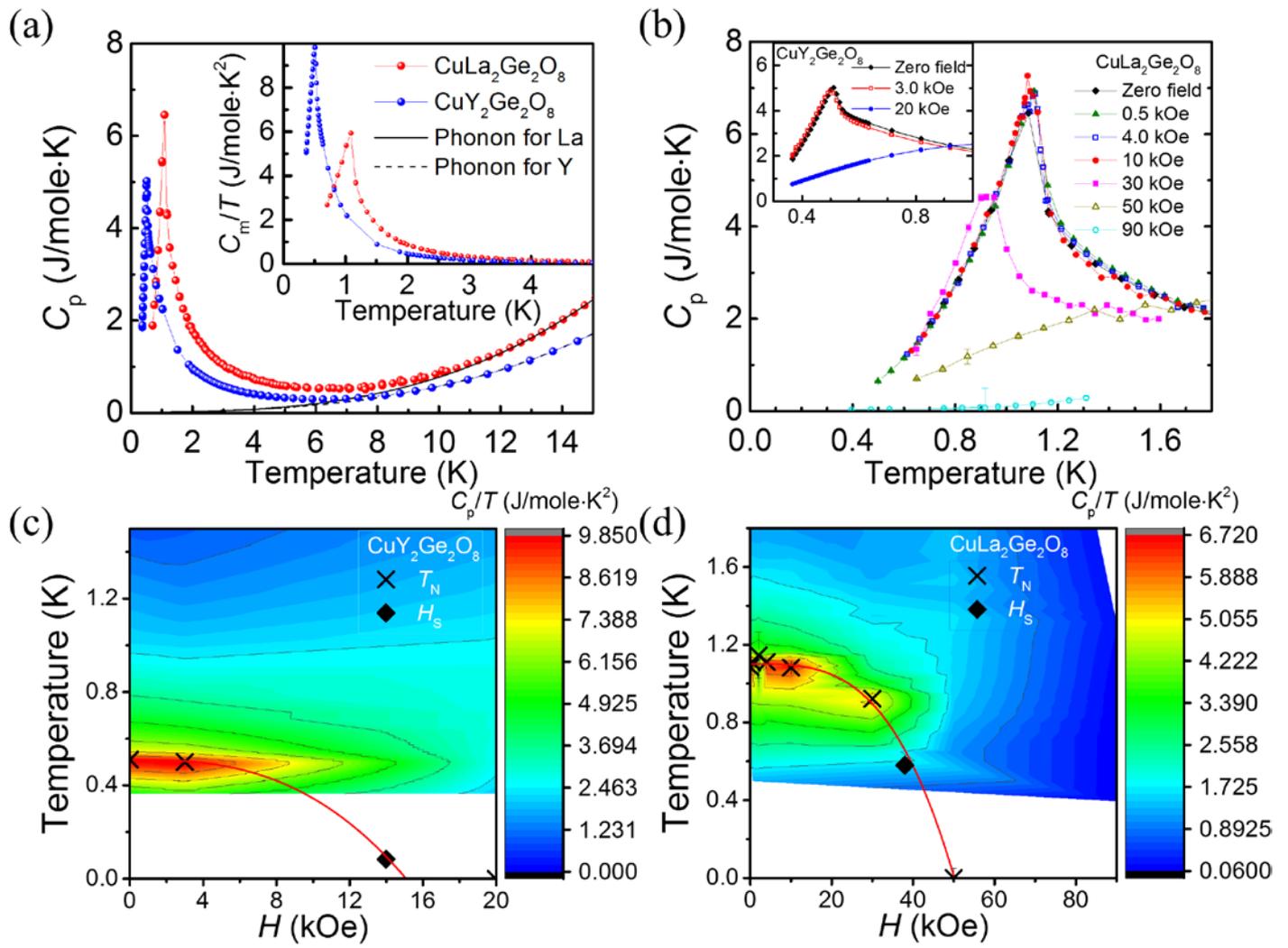

**Figure 5**

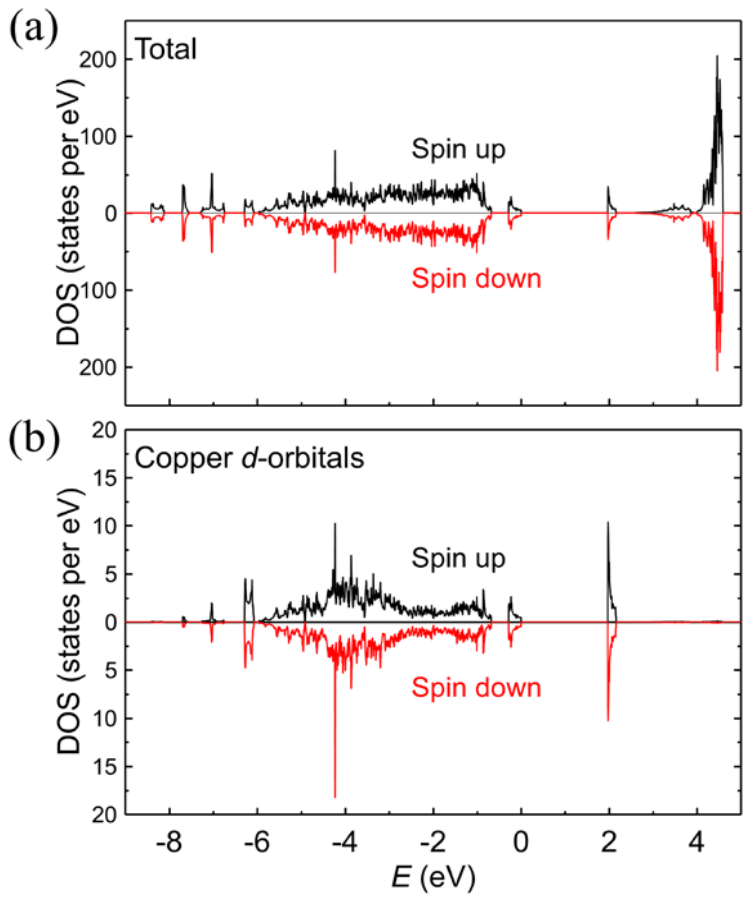

**Figure 6**

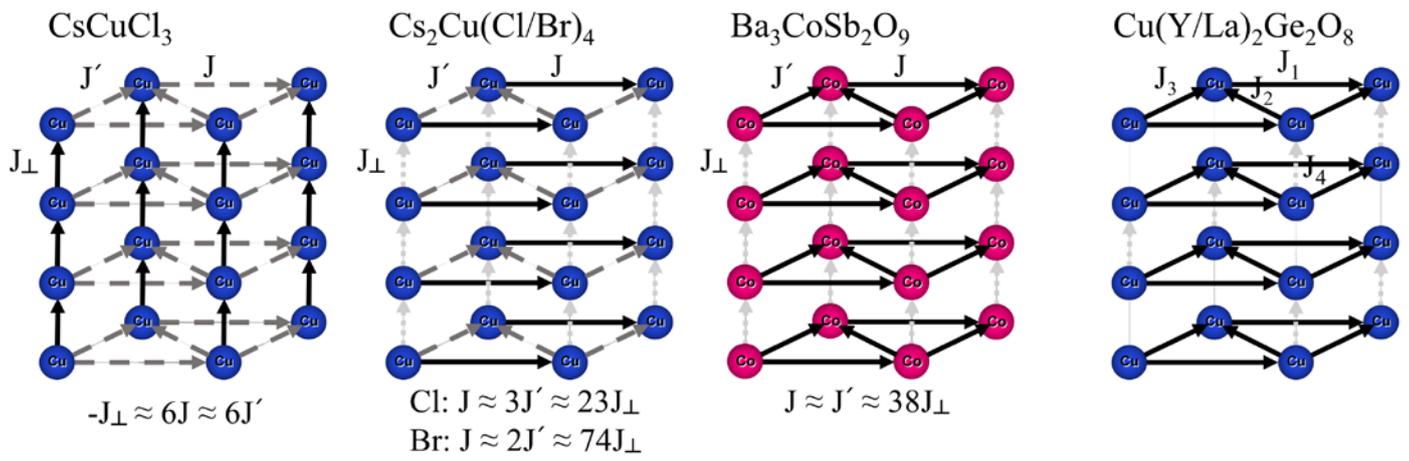

Table 1. The summary of the crystal structure of CuLa$_2$Ge$_2$O$_8$ as obtained from the single-crystal XRD refinement data shown in Figure 1(b). We used a single crystal of 50×50×100 μm$^3$.

| | Space group: $I\,1\,m\,1$ (No. 8) | | | | | |
|---|---|---|---|---|---|---|
| | Cell dimensions: $a$ (Å) = 8.587(2), $b$ (Å) = 15.565(6), $c$ (Å) = 5.199(1), $\beta$ (°) = 89.30(1), $V$ (Å$^3$) = 694.9(3) | | | | | |
| Atom | Wyckoff letter | Site symmetry | $x/a$ | $y/b$ | $z/c$ | $U_{iso}$ (Å$^2$) |
| La1 | $b$ | 1 | 0.752(1) | 0.12117(2) | 0.526(2) | 0.0094(1) |
| La2 | $b$ | 1 | 0.251(1) | 0.12038(2) | 0.526(2) | 0.0087(1) |
| Ge1 | $a$ | $m$ | 0.003(1) | 0.5000 | 0.573(2) | 0.0056(9) |
| Ge2 | $a$ | $m$ | 0.498(1) | 0.5000 | 0.479(2) | 0.006(1) |
| Ge3 | $b$ | 1 | 0.023(1) | 0.29161(7) | 0.525(2) | 0.0163(7) |
| Cu | $b$ | 1 | 0.478(1) | 0.29192(7) | 0.527(2) | 0.0051(7) |
| O1 | $a$ | $m$ | 0.8401(6) | 0.5000 | 0.778(1) | 0.0104(2) |
| O2 | $a$ | $m$ | 0.1685(6) | 0.0000 | 0.783(1) | 0.0104(2) |
| O3 | $a$ | $m$ | 0.841(2) | 0.0000 | 0.774(3) | 0.0104(2) |
| O4 | $a$ | $m$ | 0.165(2) | 0.5000 | 0.781(3) | 0.0104(2) |
| O5 | $b$ | 1 | 0.508(1) | 0.0927(5) | 0.873(2) | 0.0104(2) |
| O6 | $b$ | 1 | 0.992(1) | 0.3245(4) | 0.875(2) | 0.0104(2) |
| O7 | $b$ | 1 | 0.499(1) | 0.4069(5) | 0.680(2) | 0.0104(2) |
| O8 | $b$ | 1 | 0.999(1) | 0.1745(4) | 0.705(2) | 0.0104(2) |
| O9 | $b$ | 1 | 0.360(1) | 0.2409(3) | 0.809(2) | 0.0104(2) |
| O10 | $b$ | 1 | 0.695(1) | 0.2296(3) | 0.851(2) | 0.0104(2) |
| Atom | $U_{11}$ (Å$^2$) | $U_{22}$ (Å$^2$) | $U_{33}$ (Å$^2$) | $U_{12}$ (Å$^2$) | $U_{13}$ (Å$^2$) | $U_{23}$ (Å$^2$) |
| La1 | 0.0141(1) | 0.0049(1) | 0.0092(1) | 0.0031(4) | -0.00463(9) | 0.0013(4) |
| La2 | 0.0102(1) | 0.0049(1) | 0.0110(1) | 0.0012(4) | 0.00561(9) | -0.0000(4) |
| Ge1 | 0.0078(9) | 0.0059(7) | 0.003(1) | 0 | 0.0005(9) | 0 |
| Ge2 | 0.0062(9) | 0.0052(7) | 0.007(1) | 0 | 0.0010(9) | 0 |
| Ge3 | 0.0211(9) | 0.0130(6) | 0.0147(7) | 0.0029(7) | 0.0041(7) | -0.0011(7) |
| Cu | 0.0070(8) | 0.0015(5) | 0.0068(7) | 0.0010(6) | 0.0071(6) | -0.0011(6) |
| | Agreement factors: $\chi^2$ = 3.26, $R_{F2}$ (%) = 3.81, $R_{wF2}$ (%) = 5.91, $R_F$ (%) = 2.60 | | | | | |